\documentclass[times,twocolumn,final]{elsarticle}

\usepackage{prletters}
\usepackage{framed,multirow}
\biboptions{numbers,sort&compress}

\usepackage{multirow}
\usepackage{colortbl}
\usepackage{adjustbox}
\usepackage{stackengine}
\usepackage{url}
\usepackage{flushend}
\usepackage{xcolor, soul}

\journal{Pattern Recognition Letters}

\title{Deep Neural Networks for Automatic Speaker Recognition Do Not Learn Supra-Segmental Temporal Features}

\begin{document}

\begin{frontmatter}

\author[1]{Daniel Neururer\fnref{ec}}
\ead{neud@zhaw.ch}
\author[2]{Volker Dellwo}
\ead{volker.dellwo@uzh.ch}
\author[1,3]{Thilo Stadelmann\fnref{ec}\corref{ca}}
\ead{stdm@zhaw.ch}

\fntext[ec]{D. Neururer and T. Stadelmann contributed equally to this work.}

\affiliation[1]{organization={Centre for Artificial Intelligence, Zurich University of Applied Sciences},
addressline={Technikumstrasse 71},
postcode={8400},
city={Winterthur},
country={Switzerland}}

\affiliation[2]{organization={Department of Computational Linguistics, University of Zurich},
addressline={Andreasstrasse 15},
postcode={8050},
city={Zurich},
country={Switzerland}}

\affiliation[3]{organization={European Centre for Living Technology (ECLT)},
addressline={Ca' Bottacin, Dorsoduro 3911},
postcode={30123},
city={Venice},
country={Italy}}

\cortext[ca]{Corresponding author}

\begin{abstract}
    While deep neural networks have shown impressive results in automatic speaker recognition and related tasks, it is dissatisfactory how little is understood about what exactly is responsible for these results. Part of the success has been attributed 
    in prior work to their capability to model supra-segmental temporal information (SST), i.e., learn rhythmic-prosodic characteristics of speech in addition to spectral features. In this paper, we (i) present and apply a novel test to quantify to what extent the performance of state-of-the-art neural networks for speaker recognition can be explained by modeling SST; and (ii) present several means to force respective nets to focus more on SST and evaluate their merits. We find that a variety of CNN- and RNN-based neural network architectures for speaker recognition do not model SST to any sufficient degree, even when forced. The results provide a highly relevant basis for impactful future research into better exploitation of the full speech signal and give insights into the inner workings of such networks, enhancing explainability of deep learning for speech technologies.
\end{abstract}


\end{frontmatter}

\section{Introduction}

Deep neural networks (DNNs) have become extremely effective in speaker recognition (SR) and its sub-tasks like speaker verification (SV), identification (SI) or clustering (SC) \cite{
brown2022voxsrc}. Despite this success, deep learning remains driven by empiricism \cite{
tuggener2022enough}, and the available theoretical insights into its workings \cite{lin2017does}
all the more underline what is yet not understood about how and why DNNs arrive at such a high performance \cite{sejnowski2020unreasonable},
leaving much room for improved explainability of such models \cite{IVANOVS2021228} also to guide future research.

Meanwhile, the key to human top performance in SR (specifically, in challenging environments) is to make use of a comprehensive variety of spectro-temporal acoustic-phonetic information in speech \cite{rose2002forensic, hansen2015speaker}. Particularly, short-term spectral information, equating to frame-based acoustic information (\textbf{FBA}) in automatic systems, is supplemented in humans by supra-segmental temporal information (\textbf{SST}), also referred to as speech prosody. SST varies between individuals \cite{leemann2014speaker, SHAHNAWAZUDDIN2020213}
and is beneficial for automatic SR systems \cite{stadelmann2009unfolding}. Latter authors provided first evidence that succeeding in adequately modeling SST in addition to FBA holds the potential for an order of magnitude less SR errors. Further evidence has been provided using convolutional and recurrent deep learning architectures \cite{lukic2016speaker, lukic2017learning, stadelmann2018capturing}, claiming that the achieved gains are due to the superior sequence modeling capabilities of the DNNs that are ``successfully capturing prosodic information'' \cite{stadelmann2018capturing} -- without explicitly testing this. In a similar vein \cite{zhao2020improving, ye2021deep}: Zhao et al. \cite{zhao2020improving} argue that their BLSTM-enhanced \cite{graves2013speech} SV DNN is ``\emph{supposedly} phonetically aware'' and modeling ``context information, which \emph{could} facilitate the ResNet to [\dots] suppress the environmental variations'', \emph{because} BLSTM layers have the capability to model long ranges. Yet, these claims have never been verified. 

If DNNs would not model SST adequately (but achieve their superior results otherwise by focusing on FBA alone), this would imply that the predicted performance gains \cite{stadelmann2009unfolding} are still to be realized. More specifically, if it could be quantified to what extent state-of-the-art DNN-based SR systems actually do or do not exploit SST, this would (a) add explanation to a high-performing but opaque class of SR models, (b) show specific directions for targeted future research (concretely targeting the modelling of SST, if it turns out to be under-exploited), and (c) verify theoretical as well as empirical findings in earlier studies, practically guiding future developments. This way, the found discrepancies between DNNs theoretical capabilities and their practical workings are not just uncovered, but can be reconciled in the future. 

In this paper, we experimentally analyze this hypothesis of superior modeling of supra-segmental temporal features through contemporary DNNs made in prior work, suspecting that it happens less than assumed. Our motivation is to better understand and improve DNNs in their ability to model fundamental aspects of the speech signal inherently. Hence, we present a novel approach to quantify what amount of sequence modeling is actually occurring, and find the hypothesis of superior SST modeling has to be rejected: Current DNNs lazily rest on phone-level acoustic features alone (cf. Sec. \ref{sec_quantify}). We offer further analyses of this undesirable \emph{``deep cheating''} phenomenon (cf. Sec. \ref{sec_force}) by gradually removing the speaker-discriminant information contained in FBA features through innovative experimental setups, thus forcing models to rely on other sources of information. Still they do not switch to model available SST, as evidenced in extensive experiments using diverse state-of-the-art DNN architectures for SV using TIMIT \cite{fisher1986the} and VoxCeleb \cite{nagrani2017voxceleb} benchmarks. 

To the best of our knowledge, this paper represents the first systematic study into the modeling of SST by DNNs. Its contributions stand further out in several ways: First, our results explain what DNNs for SR do and do not model, namely that they overfit on the easily exploitable FBA to the point where they nearly ignore SST information, rectifying earlier published conjectures. Second, this opens strategic directions for future SR research that are perpendicular to other current research trends, namely to inquire into better exploitation of SST (cf. Sec. \ref{sec_conclusion}) to realize the predicted performance gain of one order of magnitude less errors \cite{stadelmann2009unfolding}. Third, the developed benchmark suite of experimental protocol, test metric and established results makes progress in this direction quantifiable.

\newcommand{\OrigSeg}{OS}
\newcommand{\ShuffleSeg}{SS}
\newcommand{\ShuffleUtterance}{SU}

\begin{figure*}[t]
  \centering
  \includegraphics[width=\textwidth]{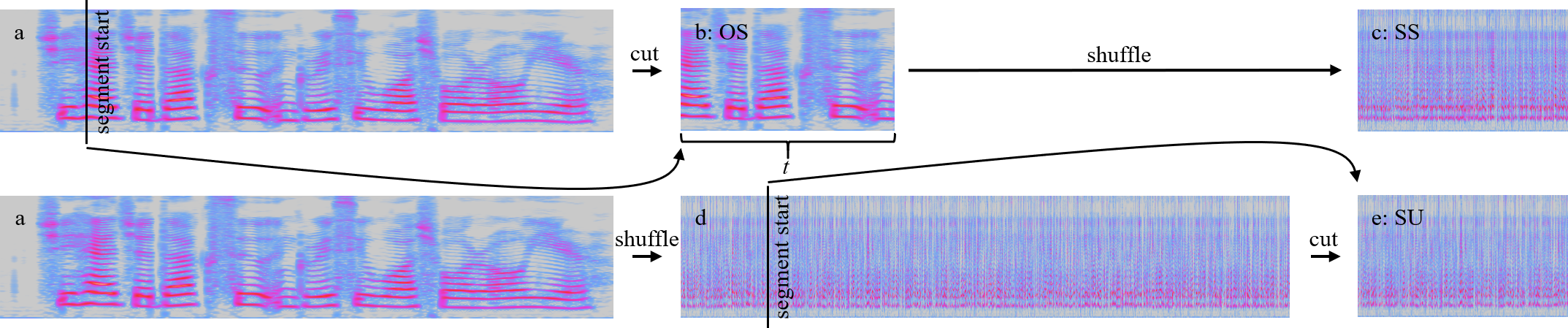}
  \caption{Segment creation with and without SST: Starting (a) from a spectrogram per varying-length \emph{utterance}, we extract \emph{segments} of fixed length $t$ from a starting point according to $3$ segment-drawing strategies as follows. Original Segment (\textbf{\OrigSeg}): just cut out (b) the respective part; Shuffled within Segment (\textbf{\ShuffleSeg}): additionally, shuffle (c) the columns of the previous output; Shuffled within Utterance (\textbf{\ShuffleUtterance}): globally shuffle all columns (d) prior to cutting (e).}
  \label{fig_segment_creation}
\end{figure*}

\section{Time scrambling approach to quantify SST exploitation}
\label{sec_quantify}

\subsection{Objective and related work}

We study SR DNNs that receive spectrogram-like input and are interested in quantifying to which extent they rely on FBA, i.e., spectral characteristics as  contained in a single frame of MFCCs \cite{davis1978evaluation}, and to which extent they exploit SST, i.e., information that is contained in the trajectory over many frames, like intonation and rhythm. Works like \cite{stadelmann2018capturing, zhao2020improving} assert that such sequence learning is happening automatically because it is a reasonable explanation for the achieved superior results, given the general sequence learning capabilities of the models. On the other hand, Soleymani et al. \cite{soleymani2018prosodic} do not rely on CNNs to pick up SST automatically just via filters that extend in time: In addition to feeding mel-spectrograms into convolutional layers to extract FBA, they feed hand-crafted prosodic features for late fusion into the DNN to achieve prosody-enhanced SV. 

The survey by Bai and Zhang \cite{bai2021speaker} shows that these two views are omnipresent in the literature: Some works use DNN architectures inspired by computer vision \cite{he2016deep, stadelmann2019beyond}, thereby asserting that exploiting dynamics will happen via filters and average pooling along the time axis in the same way that such architectures pick up image information along the x-axis automatically. According to Bai and Zhang, this is ``the most common [temporal] pooling function''. At the same time, they survey works that deal explicitly with integrating FBA over time, either via collecting statistics \cite{snyder2017deep, PINHEIRO2021100}, applying self-attention \cite{wu2020vector} or performing a trainable soft clustering of FBA \cite{xie2019utterance}. In this section, we lay the foundation for an informed decision on these options by quantifying how well DNNs \emph{inherently} capture SST in spectrogram-like input without additional explicit care for temporal dynamics.

\subsection{Methodology and experimental setup}

The authors of \cite{stadelmann2009unfolding} used a simple test to probe human reliance on SST: They randomly shuffled the columns of a spectrogram, re-synthesis the result back to audio, and conducted human SR experiments. This way, they left all FBA information intact but removed all SST within said spectrograms. Inspired by this time scrambling approach, we aim to quantify the amount of actual SST exploitation in a DNN by comparing SR performance on original input data with performance on time-scrambled input that contains no SST: Starting from a mel-spectrogram input (front-end processing similar to \cite{stadelmann2018capturing} for comparability), this is achieved by randomizing the order of speech frames, i.e., columns in the spectrogram, thereby removing all original sequence information. The anticipated drop in recognition performance would confirm an anti-proportionally good exploitation of SST through DNNs, while no drop would at least mean that the DNN can compensate missing SST equivalently with the remaining FBA information, hence does not rely on SST exploitation. 

Specifically, we consider three ways to extract fixed-length time-scrambled \emph{segments} from a variable-length \emph{utterance} (cf. Fig. \ref{fig_segment_creation}) according to the following reasoning: \textbf{\OrigSeg} segments will contain FBA and SST, offering a DNN all options to learn these lower- and higher-level features. \textbf{\ShuffleSeg} segments will, due to a random trajectory of frames, contain no speaker-specific dynamics, offering DNNs only the option to learn about FBA. \textbf{\ShuffleUtterance} segments constitute certain middle ground, biased towards FBA: While still containing no SST, they draw frames from the full utterance rather than a limited segment, offering a richer sample of FBA. 

We carried out the following initial experiments on the TIMIT database which provides a controlled acoustic recording environment and normalized utterances ($630$ speakers under clean studio conditions, $10$ sentences per speaker of on average $3$s length, $462$ speakers in the training set). This laboratory setup enabled us to first study voice recognition performance of a DNN model \emph{per se}, without complications induced through the environment (background noise, cross talk, etc.). As our main interest is in understanding and improving DNN modeling for SR (in contrast to improving a complete SR pipeline), we used three different DNN architectures to sample the space of simple to advanced models without explicit care for temporal modeling: (a) As a simple baseline, an adapted version of the vanilla convolutional network (CNN) of \cite{lukic2017learning}, sped up by a CosFace loss function \cite{wang2018cosface} instead of the computationally heavy KL divergence; (b) as the first model that claimed to model prosody, the recurrent neural network (RNN) of \cite{stadelmann2018capturing}; (c) as a recent architecture, the ResNet34s (ResNet) of \cite{xie2019utterance} that innately contains a GhostVLAD layer to aggregate FBA per segment, adapted to the experimental setup of the previous models with respect to front-end processing (see below); (d) additionally, to account for recent developments in state-of-the-art approaches, the Fast ResNet-34 (F-ResNet) of \cite{chung2020in}\footnote{Sourced from \url{https://github.com/clovaai/voxceleb_trainer}.} that also serves as a competitive baseline in the VoxSRC SR benchmarking efforts \cite{brown2022voxsrc}, 
used here with its original parameters also for front-end processing (and hence not directly comparable to the first three models). Utmost care has been taken to keep all not explicitly mentioned parameters (here or below) for, e.g., front-end processing, DNN architecture and training, embedding extraction, and classification equal to the respective original work in order to allow for comparisons. Hence, the first three models allow a comparison with prior work on modeling dynamic voice features with DNNs, while the fourth model allows a comparison with the current state-of-the art.
We expected all models to suffer performance loss in increasing magnitude when fed with shuffled frames within segments due to increasing capability to model SST.

\newcommand{\markcell}[1]{\scriptsize{#1}$\Downarrow$\footnotesize}
\setlength{\tabcolsep}{2pt} 

\begin{table}[t]
    \centering
    \caption{SC results on TIMIT [MR $\mu/\sigma$]. Bold font indicates best results per model, cell coloring scales with quality per model. Cells marked by $\Downarrow$ are discussed in the text.}    
    	\begin{tabular}{llrrr}
    		\hline
    		 \multicolumn{2}{c}{\scriptsize{$\downarrow$ training / test $\rightarrow$}}
    		 & \multicolumn{1}{c}{\OrigSeg}
    		 & \multicolumn{1}{c}{\ShuffleUtterance}
    		 & \multicolumn{1}{c}{\ShuffleSeg} \\
    
    		\hline
    		\multirow{3}{*}{CNN \cite{lukic2017learning}}
    		
    		 & \OrigSeg
    		 & \cellcolor[RGB]{  0, 255, 0} \markcell{a} \textbf{0.00 \scriptsize{$\sigma$0.00}}
    		 & \cellcolor[RGB]{255,   0, 0} 9.75 \scriptsize{$\sigma$0.94}
    		 & \cellcolor[RGB]{255,  39, 0} \markcell{a,f} 9.00 \scriptsize{$\sigma$2.15} \\
    
    		 & \ShuffleUtterance
    		 & \cellcolor[RGB]{255,  65, 0} 8.50 \scriptsize{$\sigma$2.42}
    		 & \cellcolor[RGB]{ 26, 255, 0} \markcell{e} 0.50 \scriptsize{$\sigma$0.61}
    		 & \cellcolor[RGB]{ 91, 255, 0}1.75 \scriptsize{$\sigma$0.61} \\
    
    		 & \ShuffleSeg
    		 & \cellcolor[RGB]{255,  39, 0} \markcell{a} 9.00 \scriptsize{$\sigma$1.66}
    		 & \cellcolor[RGB]{ 52, 255, 0}1.00 \scriptsize{$\sigma$0.50}
    		 & \cellcolor[RGB]{ 65, 255, 0} \markcell{b} 1.25 \scriptsize{$\sigma$0.00} \\
    
    		\hline
    		\multirow{3}{*}{RNN \cite{stadelmann2018capturing}}
    		
    		 & \OrigSeg
    		 & \cellcolor[RGB]{170, 255, 0} \markcell{d} 1.25 \scriptsize{$\sigma$1.12}
    		 & \cellcolor[RGB]{255, 136, 0} 2.75 \scriptsize{$\sigma$0.94}
    		 & \cellcolor[RGB]{255, 135, 0} \markcell{f} 2.75 \scriptsize{$\sigma$0.50} \\
    
    		 & \ShuffleUtterance
    		 & \cellcolor[RGB]{255,   0, 0} 3.75 \scriptsize{$\sigma$1.37}
    		 & \cellcolor[RGB]{  0, 255, 0} \textbf{0.00 \scriptsize{$\sigma$0.00}}
    		 & \cellcolor[RGB]{255, 169, 0} 2.50 \scriptsize{$\sigma$1.58} \\
    
    		 & \ShuffleSeg
    		 & \cellcolor[RGB]{255, 238, 0} 2.00 \scriptsize{$\sigma$1.00}
    		 & \cellcolor[RGB]{170, 255, 0} 1.25 \scriptsize{$\sigma$0.79}
    		 & \cellcolor[RGB]{ 33, 255, 0} \markcell{d} 0.25 \scriptsize{$\sigma$0.50} \\
    
    		\hline
    		\multirow{3}{*}{ResNet \cite{xie2019utterance}}
    		
    		 & \OrigSeg
    		 & \cellcolor[RGB]{  0, 255, 0} \markcell{d} \textbf{1.00 \scriptsize{$\sigma$0.94}}
    		 & \cellcolor[RGB]{255, 157, 0} 8.25 \scriptsize{$\sigma$4.78}
    		 & \cellcolor[RGB]{255,   0, 0} \markcell{f} 11.50 \scriptsize{$\sigma$4.29} \\
    
    		 & \ShuffleUtterance
    		 & \cellcolor[RGB]{ 72, 255, 0} 2.50 \scriptsize{$\sigma$1.77}
    		 & \cellcolor[RGB]{  0, 255, 0} \textbf{1.00 \scriptsize{$\sigma$0.50}}
    		 & \cellcolor[RGB]{ 97, 255, 0} 3.00 \scriptsize{$\sigma$1.27} \\
    
    		 & \ShuffleSeg
    		 & \cellcolor[RGB]{ 84, 255, 0} 2.75 \scriptsize{$\sigma$0.94}
    		 & \cellcolor[RGB]{ 12, 255, 0} 1.25 \scriptsize{$\sigma$1.12}
    		 & \cellcolor[RGB]{  0, 255, 0} \markcell{d} \textbf{1.00 \scriptsize{$\sigma$0.94}} \\

    		\hline
    		\multirow{3}{*}{F-ResNet \cite{chung2020in}}
    		
       		 & \OrigSeg
       		 & \cellcolor[RGB]{117, 255,   0} 11.50 \scriptsize{$\sigma$2.15}
       		 & \cellcolor[RGB]{255,   0,   0} 37.50 \scriptsize{$\sigma$4.18}
       		 & \cellcolor[RGB]{255,  57,   0} 33.75 \scriptsize{$\sigma$4.18} \\

   		     & \ShuffleUtterance
   		     & \cellcolor[RGB]{192, 255,   0} 16.50 \scriptsize{$\sigma$2.42}
   		     & \cellcolor[RGB]{ 30, 255,   0}  5.75 \scriptsize{$\sigma$1.70}
       		 & \cellcolor[RGB]{  7, 255,   0}  4.25 \scriptsize{$\sigma$1.00} \\

   		     & \ShuffleSeg
   		     & \cellcolor[RGB]{178, 255,   0} 15.50 \scriptsize{$\sigma$2.57}
   		     & \cellcolor[RGB]{ 45, 255,   0}  6.75 \scriptsize{$\sigma$1.50}
   		     & \cellcolor[RGB]{  0, 255,   0} \markcell{j} \textbf{ 3.75 \scriptsize{$\sigma$0.79}} \\
    
    		\hline
    	\end{tabular}
    \label{tab_timit_clustering}
\end{table}

\begin{table}[!t]
    \centering
    \caption{SV results on TIMIT [EER $\mu/\sigma$]. As with the SC results on TIMIT in Tab. \ref{tab_timit_clustering}, the F-ResNet is out of competition here due to different front-end processing.}    
    	\begin{tabular}{llrrr}
    		\hline
    		 \multicolumn{2}{c}{\scriptsize{$\downarrow$ training / test $\rightarrow$}}
    		 & \multicolumn{1}{c}{\OrigSeg}
    		 & \multicolumn{1}{c}{\ShuffleUtterance}
    		 & \multicolumn{1}{c}{\ShuffleSeg} \\

    		\hline
    		\multirow{3}{*}{CNN \cite{lukic2017learning}}
    		
    		 & \OrigSeg
    		 & \cellcolor[RGB]{ 80, 255, 0} \markcell{c} 6.38 \scriptsize{$\sigma$0.12}
    		 & \cellcolor[RGB]{255,   0, 0} 12.02 \scriptsize{$\sigma$0.51}
    		 & \cellcolor[RGB]{255,   9, 0} \markcell{f} 11.90 \scriptsize{$\sigma$0.46} \\
    		
    		 & \ShuffleUtterance
    		 & \cellcolor[RGB]{245, 255, 0}8.55 \scriptsize{$\sigma$0.49}
    		 & \cellcolor[RGB]{ 16, 255, 0} \markcell{g-i} 5.55 \scriptsize{$\sigma$0.06}
    		 & \cellcolor[RGB]{ 60, 255, 0} 6.12 \scriptsize{$\sigma$0.12} \\
    		
    		 & \ShuffleSeg
    		 & \cellcolor[RGB]{215, 255, 0} 8.16 \scriptsize{$\sigma$0.42}
    		 & \cellcolor[RGB]{  0, 255, 0} \textbf{5.33 \scriptsize{$\sigma$0.18}}
    		 & \cellcolor[RGB]{ 34, 255, 0} \markcell{c} 5.78 \scriptsize{$\sigma$0.16} \\

    		\hline
    		\multirow{3}{*}{RNN \cite{stadelmann2018capturing}}
    		
    		 & \OrigSeg
    		 & \cellcolor[RGB]{  0, 255, 0} \markcell{d} \textbf{3.53 \scriptsize{$\sigma$0.07}}
    		 & \cellcolor[RGB]{255,   0, 0} 4.19 \scriptsize{$\sigma$0.09}
    		 & \cellcolor[RGB]{255, 224, 0} \markcell{f} 3.90 \scriptsize{$\sigma$0.12} \\

    		 & \ShuffleUtterance
    		 & \cellcolor[RGB]{255, 149, 0} 3.99 \scriptsize{$\sigma$0.16}
    		 & \cellcolor[RGB]{189, 255, 0} 3.78 \scriptsize{$\sigma$0.10}
    		 & \cellcolor[RGB]{ 97, 255, 0} 3.66 \scriptsize{$\sigma$0.13} \\

    		 & \ShuffleSeg
    		 & \cellcolor[RGB]{255, 144, 0} 4.00 \scriptsize{$\sigma$0.07}
    		 & \cellcolor[RGB]{255, 232, 0} 3.89 \scriptsize{$\sigma$0.06}
    		 & \cellcolor[RGB]{  4, 255, 0} \markcell{d,g-i} 3.54 \scriptsize{$\sigma$0.05} \\

    		\hline
    		\multirow{3}{*}{ResNet \cite{xie2019utterance}}
    		
    		 & \OrigSeg
    		 & \cellcolor[RGB]{  0, 255, 0} \textbf{4.96 \scriptsize{$\sigma$0.19}}
    		 & \cellcolor[RGB]{255,   0, 0} 10.34 \scriptsize{$\sigma$1.56}
    		 & \cellcolor[RGB]{255, 106, 0} \markcell{f} 9.21 \scriptsize{$\sigma$1.15} \\

    		 & \ShuffleUtterance
    		 & \cellcolor[RGB]{154, 255, 0} 6.59 \scriptsize{$\sigma$0.25}
    		 & \cellcolor[RGB]{122, 255, 0} 6.25 \scriptsize{$\sigma$0.23}
    		 & \cellcolor[RGB]{133, 255, 0} 6.37 \scriptsize{$\sigma$0.35} \\

    		 & \ShuffleSeg
    		 & \cellcolor[RGB]{ 87, 255, 0} 5.89 \scriptsize{$\sigma$0.25}
    		 & \cellcolor[RGB]{108, 255, 0} 6.11 \scriptsize{$\sigma$0.31}
    		 & \cellcolor[RGB]{ 79, 255, 0} \markcell{g-i} 5.80 \scriptsize{$\sigma$0.11} \\

    		\hline
    		\multirow{3}{*}{F-ResNet \cite{chung2020in}}
    		
    		 & \OrigSeg
       		 & \cellcolor[RGB]{114, 255,   0} 12.20 \scriptsize{$\sigma$0.25}
       		 & \cellcolor[RGB]{255,   0,   0} 23.41 \scriptsize{$\sigma$1.73}
       		 & \cellcolor[RGB]{255, 104,   0} 20.47 \scriptsize{$\sigma$1.50} \\

       		 & \ShuffleUtterance
       		 & \cellcolor[RGB]{217, 255,   0} 15.12 \scriptsize{$\sigma$0.84}
       		 & \cellcolor[RGB]{ 53, 255,   0} 10.46 \scriptsize{$\sigma$0.28}
       		 & \cellcolor[RGB]{ 26, 255,   0}  9.69 \scriptsize{$\sigma$0.20} \\

       		 & \ShuffleSeg
       		 & \cellcolor[RGB]{246, 255,   0} 15.91 \scriptsize{$\sigma$0.90}
       		 & \cellcolor[RGB]{ 32, 255,   0}  9.86 \scriptsize{$\sigma$0.11}
       		 & \cellcolor[RGB]{  0, 255,   0} \markcell{j} \textbf{ 8.95 \scriptsize{$\sigma$0.13}} \\

    		\hline
    	\end{tabular}
    \label{tab_timit_verification}
\end{table}

For training of the CNN, RNN, and ResNet model, to warrant backward comparability with prior work, in each of $128$ epochs we draw one $t=1$s long segment (to account for short utterances \cite{stadelmann2010dimension}) from a random location per utterance in the training set, using a batch size of $100$ and otherwise a similar experimental setup as \cite{stadelmann2018capturing}. For the F-ResNet, to keep forward comparability with current SR benchmarking efforts, we keep all hyperparameters similar to the experimental setup in \cite{chung2020in}, specifically using $t=2$s long training segments ($t=4$s for evaluation), $500$ epochs and a batch size of $800$.
We repeat this for each of the \OrigSeg/\ShuffleSeg/\ShuffleUtterance\ strategies to draw segments with/without SST, thus training three separate versions of each model. We use these models to extract embeddings for two downstream tasks: For SV, we pair every sentence in the TIMIT test set with every other sentence therein, and evaluate the equal error rate (EER) as the standard metric used for SV \cite{brown2022voxsrc}, resulting in $2.82$ mio. pairwise comparisons. For SC, we perform hierarchical clustering of $2$ utterances (comprising $2$ and $8$ concatenated sentences) per $40$ speakers as in \cite{stadelmann2018capturing}, measured by the misclassification rate (MR) as the established measure \cite{lukic2016speaker} ($6,400$ pairwise comparisons). For each reported EER/MR, we average over $5$ train/test runs and report mean and SD. 

\subsection{Results and discussion}
Results are shown in Tabs. \ref{tab_timit_clustering} and \ref{tab_timit_verification}. For the analysis, we first focus on SC results (Tab. \ref{tab_timit_clustering}) and original timing vs. segment-wise shuffled (rows/columns marked with \OrigSeg/\ShuffleSeg), ignoring the rest. With reference to the lowercase letters in the cells marked by $\Downarrow$, we can say: (a) Looking at the CNN model only, everything appears as expected as best results ($MR=0$ in all $5$ runs, the best result ever reported on this benchmark) are achieved for training and testing with \OrigSeg, and worst results are achieved when \ShuffleSeg\ is used for training \emph{or} evaluation. (b) But already looking at training \emph{and} testing using \ShuffleSeg\ raises doubts w.r.t. actual SST exploitation: $MR=1.25$ is a very good result for the task, still outperforming the previous state of the art \cite{stadelmann2018capturing}, but not using any SST (as it has been removed from the data). (c) Taking, secondly, SV results (Tab. \ref{tab_timit_verification}) into account, these doubts are confirmed. For the CNN model, training/testing using \ShuffleSeg/\ShuffleSeg\ outperforms \OrigSeg/\OrigSeg\ by $0.6$ difference in EER. (d) A similar picture is seen for the RNN on both tasks and the ResNet on SC, where \ShuffleSeg/\ShuffleSeg\ either outperforms or (within $\sigma$) equals \OrigSeg/\OrigSeg. (e) When adding \ShuffleUtterance\ for training and/or testing into the picture, this tendency is confirmed: \ShuffleUtterance/\ShuffleUtterance\ outperforms \ShuffleSeg/\ShuffleSeg\ and is almost on par with \OrigSeg/\OrigSeg\ (cf. a). (f) However, models do pick up something about inter-frame relationships as in cross conditions like \OrigSeg\ training and \ShuffleSeg\ testing, a performance drop is evident for the CNN and ResNet (not so for the RNN). We argue that this is rather the effect of mismatched train/test conditions as is usual in any machine learned model; previous discussion (specifically, that best results are achieved using \ShuffleUtterance\ in several cases) has shown that (almost) nothing useful is extracted from the trajectory. 
(j) The F-ResNet, which uses the front-end processing and architectural parameters optimized for SV on the by orders of magnitude larger VoxCeleb dataset, has expected difficulties with the tiny TIMIT benchmark (which could be cured by proper pre-training, but would not help the purpose of these experiments). It hence ran out of competition here, but will play a major role in later experiments (see Section \ref{sec_force}). For now, it suffices to say that it behaved quite similar to the ResNet. That the best scores are achieved using the \ShuffleSeg/\ShuffleSeg\ setup on both tasks can be explained with the general lack of sufficient amounts of training data, specifically when using longer segments, and that in this regime SS provides most FBA information per segment (i.e., the higher sample efficiency can partially make up for lacking training data).

Summarizing, we found that completely randomizing the order of speech frames in segments used for training and evaluation still produced state-of-the-art or even better SV (e.g., \ShuffleUtterance/\ShuffleUtterance\ for the RNN) and SC (e.g., \ShuffleSeg/\ShuffleSeg\ for the RNN) results. This is notwithstanding the obviously also state-of-the-art results of the \OrigSeg/\OrigSeg\ setup for several models (e.g., the CNN for SC; ResNet for SV) -- the point here is not that the results using \ShuffleSeg\ or \ShuffleUtterance\ are in any way superior to \OrigSeg, but that they are sometimes not worse. Given this analysis, we believe to have strong evidence that the tested DNN models do not rely on SST (they might model them, but are ultimately able to produce similar results without them). We conjecture that this is due to purely modeling FBA is easy and efficient enough: It quickly brings the loss down to a local minimum from which the training cannot recover as switching features to a different set would introduce too high intermediate losses again. This is in accordance with studies claiming that DNNs are lazy in learning complex concepts when easy ones work already \cite{mansour2019deep}, and tend to learn shortcuts \cite{degrave2021ai}. To put it drastically, DNNs ``cheat'' by ignoring the harder task of modeling dynamics if they can.

Technically speaking, the model overfits to FBA information to the point of neglecting other evidence. An ablation study gives further evidence of this: Remarkably, replacing the CNN in the TIMIT SV task with a simplistic model per speaker that only holds a row-wise average of all the speaker's training spectrograms (i.e., a single-Gaussian model with unit variance) outperforms the CNN. Apparently, having access to the full frequency resolution of the spectrogram at decision time is a larger advantage to the simple model than pooling in frequency \emph{and time} is for the CNN. Overcoming this problem calls for a novel form of regularization on the task-level, which will be explored in the following section. 

\begin{table*}[ht]
	\centering
	\caption{SV results on VoxCeleb (left) and noise-vocoded (middle) as well as resynthesized (right) TIMIT [EER $\mu/\sigma$ of $5$ runs].}
    \begin{adjustbox}{width=\textwidth}
    	\begin{tabular}{ll|rrr|rrr|rrr}
    		\hline
    		 & 
    		 & \multicolumn{3}{c|}{VoxCeleb}
    		 & \multicolumn{3}{c|}{TIMIT-NV}
    		 & \multicolumn{3}{c}{TIMIT-Syn} \\
    		
    		 \multicolumn{2}{c|}{\scriptsize{$\downarrow$ training / test $\rightarrow$}}
    		 & \multicolumn{1}{c}{\OrigSeg}
    		 & \multicolumn{1}{c}{\ShuffleUtterance}
    		 & \multicolumn{1}{c|}{\ShuffleSeg}
    		 & \multicolumn{1}{c}{\OrigSeg}
    		 & \multicolumn{1}{c}{\ShuffleUtterance}
    		 & \multicolumn{1}{c|}{\ShuffleSeg}
    		 & \multicolumn{1}{c}{\OrigSeg}
    		 & \multicolumn{1}{c}{\ShuffleUtterance}
    		 & \multicolumn{1}{c}{\ShuffleSeg} \\

            \hline
    		\multirow{3}{*}{CNN \cite{lukic2017learning}}
    		
    		 & \OrigSeg
    		 & \cellcolor[RGB]{  0, 255, 0} \markcell{g} \textbf{25.75 \scriptsize{$\sigma$0.13}}
    		 & \cellcolor[RGB]{255,   0, 0} 37.23 \scriptsize{$\sigma$0.74}
    		 & \cellcolor[RGB]{255,  12, 0} 36.96 \scriptsize{$\sigma$0.78}
    		 & \cellcolor[RGB]{220, 255, 0}  32.56 \scriptsize{$\sigma$0.62}
    		 & \cellcolor[RGB]{255,   8, 0}  35.32 \scriptsize{$\sigma$0.46}
    		 & \cellcolor[RGB]{255,   0, 0}  35.41 \scriptsize{$\sigma$0.55}
    		 & \cellcolor[RGB]{ 90, 255, 0}  46.24 \scriptsize{$\sigma$0.18}
    		 & \cellcolor[RGB]{255,   4, 0}  48.94 \scriptsize{$\sigma$0.15}
    		 & \cellcolor[RGB]{255,   0, 0}  48.97 \scriptsize{$\sigma$0.23} \\

    		 & \ShuffleUtterance
    		 & \cellcolor[RGB]{255, 201, 0} 32.70 \scriptsize{$\sigma$0.34}
    		 & \cellcolor[RGB]{ 57, 255, 0} 27.04 \scriptsize{$\sigma$0.34}
    		 & \cellcolor[RGB]{ 99, 255, 0} 27.99 \scriptsize{$\sigma$0.30}
    		 & \cellcolor[RGB]{255,  25, 0}  35.16 \scriptsize{$\sigma$0.52}
    		 & \cellcolor[RGB]{  0, 255, 0} \markcell{h} \textbf{30.39 \scriptsize{$\sigma$0.30}}
    		 & \cellcolor[RGB]{ 53, 255, 0}  30.91 \scriptsize{$\sigma$0.47}
    		 & \cellcolor[RGB]{246, 255, 0}  47.26 \scriptsize{$\sigma$0.15}
    		 & \cellcolor[RGB]{ 49, 255, 0}  45.98 \scriptsize{$\sigma$0.34}
    		 & \cellcolor[RGB]{ 77, 255, 0}  46.16 \scriptsize{$\sigma$0.27} \\
    		
    		 & \ShuffleSeg
    		 & \cellcolor[RGB]{255, 176, 0} 33.26 \scriptsize{$\sigma$0.29}
    		 & \cellcolor[RGB]{ 95, 255, 0} 27.91 \scriptsize{$\sigma$0.32}
    		 & \cellcolor[RGB]{122, 255, 0} 28.50 \scriptsize{$\sigma$0.28}
    		 & \cellcolor[RGB]{255,  15, 0}  35.25 \scriptsize{$\sigma$0.69}
    		 & \cellcolor[RGB]{ 25, 255, 0}  30.63 \scriptsize{$\sigma$0.38}
    		 & \cellcolor[RGB]{ 86, 255, 0}  31.23 \scriptsize{$\sigma$0.27}
    		 & \cellcolor[RGB]{228, 255, 0}  47.14 \scriptsize{$\sigma$0.22}
    		 & \cellcolor[RGB]{ 33, 255, 0}  45.88 \scriptsize{$\sigma$0.12}
    		 & \cellcolor[RGB]{  0, 255, 0} \markcell{i} \textbf{45.66 \scriptsize{$\sigma$0.12}} \\

            \hline
    		\multirow{3}{*}{RNN \cite{stadelmann2018capturing}}
    		
    		 & \OrigSeg
    		 & \cellcolor[RGB]{  0, 255, 0} \markcell{g} \textbf{20.67 \scriptsize{$\sigma$0.23}}
    		 & \cellcolor[RGB]{255,   0, 0} 30.67 \scriptsize{$\sigma$0.36}
    		 & \cellcolor[RGB]{255,  33, 0} 30.00 \scriptsize{$\sigma$0.32}
    		 & \cellcolor[RGB]{  0, 255, 0} \markcell{h} \textbf{19.34 \scriptsize{$\sigma$0.16}}
    		 & \cellcolor[RGB]{255,   0, 0}  27.20 \scriptsize{$\sigma$0.42}
    		 & \cellcolor[RGB]{255,  70, 0}  26.12 \scriptsize{$\sigma$0.44}
    		 & \cellcolor[RGB]{  0, 255, 0} \markcell{i} \textbf{40.39 \scriptsize{$\sigma$0.07}}
    		 & \cellcolor[RGB]{255,   0, 0}  44.29 \scriptsize{$\sigma$0.65}
    		 & \cellcolor[RGB]{255, 242, 0}  42.43 \scriptsize{$\sigma$1.40} \\

    		 & \ShuffleUtterance
    		 & \cellcolor[RGB]{255, 227, 0} 26.20 \scriptsize{$\sigma$0.18}
    		 & \cellcolor[RGB]{ 68, 255, 0} 22.02 \scriptsize{$\sigma$0.10}
    		 & \cellcolor[RGB]{147, 255, 0} 23.57 \scriptsize{$\sigma$0.09}
    		 & \cellcolor[RGB]{234, 255, 0}  22.95 \scriptsize{$\sigma$0.24}
    		 & \cellcolor[RGB]{138, 255, 0}  21.48 \scriptsize{$\sigma$0.40}
    		 & \cellcolor[RGB]{117, 255, 0}  21.15 \scriptsize{$\sigma$0.25}
    		 & \cellcolor[RGB]{255,  86, 0}  43.63 \scriptsize{$\sigma$0.35}
    		 & \cellcolor[RGB]{201, 255, 0}  41.93 \scriptsize{$\sigma$0.26}
    		 & \cellcolor[RGB]{163, 255, 0}  41.64 \scriptsize{$\sigma$0.25} \\
    
    		 & \ShuffleSeg
    		 & \cellcolor[RGB]{255, 121, 0} 28.28 \scriptsize{$\sigma$1.30}
    		 & \cellcolor[RGB]{255, 222, 0} 26.30 \scriptsize{$\sigma$0.59}
    		 & \cellcolor[RGB]{255, 208, 0} 26.58 \scriptsize{$\sigma$0.84}
    		 & \cellcolor[RGB]{225, 255, 0}  22.82 \scriptsize{$\sigma$0.40}
    		 & \cellcolor[RGB]{165, 255, 0}  21.89 \scriptsize{$\sigma$0.25}
    		 & \cellcolor[RGB]{109, 255, 0}  21.04 \scriptsize{$\sigma$0.12}
    		 & \cellcolor[RGB]{255,  88, 0}  43.62 \scriptsize{$\sigma$0.21}
    		 & \cellcolor[RGB]{255, 227, 0}  42.55 \scriptsize{$\sigma$0.34}
    		 & \cellcolor[RGB]{149, 255, 0}  41.53 \scriptsize{$\sigma$0.23} \\

    		 \hline
    		 \multirow{3}{*}{ResNet \cite{xie2019utterance}}
    		
    		 & \OrigSeg
    		 & \cellcolor[RGB]{  0, 255, 0} \markcell{g} \textbf{12.49 \scriptsize{$\sigma$0.15}}
    		 & \cellcolor[RGB]{255,   0, 0} 34.11 \scriptsize{$\sigma$0.54}
    		 & \cellcolor[RGB]{255,  45, 0} 32.19 \scriptsize{$\sigma$0.39}
    		 & \cellcolor[RGB]{  0, 255, 0} \markcell{h} \textbf{21.12 \scriptsize{$\sigma$0.43}}
    		 & \cellcolor[RGB]{255,   0, 0}  37.83 \scriptsize{$\sigma$1.17}
    		 & \cellcolor[RGB]{255,  38, 0}  36.57 \scriptsize{$\sigma$1.45}
    		 & \cellcolor[RGB]{  0, 255, 0} \markcell{i} \textbf{40.33 \scriptsize{$\sigma$1.32}}
    		 & \cellcolor[RGB]{255,   0, 0}  47.28 \scriptsize{$\sigma$2.06}
    		 & \cellcolor[RGB]{255,  49, 0}  46.60 \scriptsize{$\sigma$2.02} \\

    		 & \ShuffleUtterance
    		 & \cellcolor[RGB]{225, 255, 0} 22.05 \scriptsize{$\sigma$0.43}
    		 & \cellcolor[RGB]{155, 255, 0} 19.08 \scriptsize{$\sigma$0.26}
    		 & \cellcolor[RGB]{177, 255, 0} 20.02 \scriptsize{$\sigma$0.16}
    		 & \cellcolor[RGB]{180, 255, 0}  27.03 \scriptsize{$\sigma$0.63}
    		 & \cellcolor[RGB]{ 68, 255, 0}  23.38 \scriptsize{$\sigma$0.41}
    		 & \cellcolor[RGB]{ 88, 255, 0}  24.02 \scriptsize{$\sigma$0.25}
    		 & \cellcolor[RGB]{228, 255, 0}  43.44 \scriptsize{$\sigma$0.86}
    		 & \cellcolor[RGB]{194, 255, 0}  42.97 \scriptsize{$\sigma$0.51}
    		 & \cellcolor[RGB]{170, 255, 0}  42.65 \scriptsize{$\sigma$0.59} \\
    
    		 & \ShuffleSeg
    		 & \cellcolor[RGB]{194, 255, 0} 20.74 \scriptsize{$\sigma$0.46}
    		 & \cellcolor[RGB]{201, 255, 0} 21.02 \scriptsize{$\sigma$0.34}
    		 & \cellcolor[RGB]{185, 255, 0} 20.36 \scriptsize{$\sigma$0.23}
    		 & \cellcolor[RGB]{187, 255, 0}  27.25 \scriptsize{$\sigma$1.37}
    		 & \cellcolor[RGB]{ 74, 255, 0}  23.57 \scriptsize{$\sigma$0.46}
    		 & \cellcolor[RGB]{ 67, 255, 0}  23.32 \scriptsize{$\sigma$0.58}
    		 & \cellcolor[RGB]{157, 255, 0}  42.48 \scriptsize{$\sigma$0.45}
    		 & \cellcolor[RGB]{201, 255, 0}  43.07 \scriptsize{$\sigma$0.72}
    		 & \cellcolor[RGB]{ 92, 255, 0}  41.59 \scriptsize{$\sigma$0.36} \\
    
            \hline
    		 \multirow{3}{*}{F-ResNet \cite{chung2020in}}
    		 
    		 & \OrigSeg
       		 & \cellcolor[RGB]{  0, 255,   0} \markcell{g} \textbf{ 2.39 \scriptsize{$\sigma$0.05}}
       		 & \cellcolor[RGB]{255,   0,   0} 25.02 \scriptsize{$\sigma$1.21}
       		 & \cellcolor[RGB]{255,  35,   0} 23.45 \scriptsize{$\sigma$1.29}
    		 & \cellcolor[RGB]{111, 255,   0} \markcell{k} 24.72 \scriptsize{$\sigma$0.54}
       		 & \cellcolor[RGB]{255,   0,   0} 37.17 \scriptsize{$\sigma$0.57}
   	    	 & \cellcolor[RGB]{255,  84,   0} 34.54 \scriptsize{$\sigma$0.66}
    		 & \cellcolor[RGB]{  0, 255,   0} \markcell{k} \textbf{39.06 \scriptsize{$\sigma$0.50}}
   		     & \cellcolor[RGB]{255,   0,   0} 47.52 \scriptsize{$\sigma$0.58}
   		     & \cellcolor[RGB]{255,  62,   0} 46.48 \scriptsize{$\sigma$1.02} \\

    		 & \ShuffleUtterance
       		 & \cellcolor[RGB]{194, 255,   0} 11.00 \scriptsize{$\sigma$0.57}
       		 & \cellcolor[RGB]{ 97, 255,   0}  6.69 \scriptsize{$\sigma$0.11}
       		 & \cellcolor[RGB]{ 94, 255,   0}  6.58 \scriptsize{$\sigma$0.15} 
       		 & \cellcolor[RGB]{255, 183,   0} 31.46 \scriptsize{$\sigma$0.86}
       		 & \cellcolor[RGB]{ 37, 255,   0} 22.40 \scriptsize{$\sigma$0.42}
       		 & \cellcolor[RGB]{ 14, 255,   0} 21.67 \scriptsize{$\sigma$0.49}   		     
    		 & \cellcolor[RGB]{228, 255,   0} 42.85 \scriptsize{$\sigma$0.47}
       		 & \cellcolor[RGB]{ 62, 255,   0} 40.10 \scriptsize{$\sigma$0.11}
       		 & \cellcolor[RGB]{ 68, 255,   0} 40.19 \scriptsize{$\sigma$0.10} \\
    
    		 & \ShuffleSeg
             & \cellcolor[RGB]{190, 255,   0} 10.85 \scriptsize{$\sigma$0.43}
       		 & \cellcolor[RGB]{104, 255,   0}  7.02 \scriptsize{$\sigma$0.24}
       		 & \cellcolor[RGB]{ 95, 255,   0}  6.60 \scriptsize{$\sigma$0.19}     		 
             & \cellcolor[RGB]{255, 183,   0} 31.44 \scriptsize{$\sigma$1.11}
       		 & \cellcolor[RGB]{ 26, 255,   0} 22.04 \scriptsize{$\sigma$0.12}
       		 & \cellcolor[RGB]{  0, 255,   0} \markcell{k} \textbf{21.24 \scriptsize{$\sigma$0.17}}    		 
    		 & \cellcolor[RGB]{252, 255,   0} 43.23 \scriptsize{$\sigma$0.29}
       		 & \cellcolor[RGB]{ 83, 255,   0} 40.44 \scriptsize{$\sigma$0.18}
       		 & \cellcolor[RGB]{ 78, 255,   0} 40.36 \scriptsize{$\sigma$0.20} \\
       		 
    		\hline
    	\end{tabular}
    \end{adjustbox}
	\label{tab_forcing}    
\end{table*}

\section{Regularization approaches to increase SST exploitation}
\label{sec_force}

\subsection{Objective and context}

Our objective here is to find ways to force models to revert to SST. As one implication of the results of Sec. \ref{sec_quantify} is that the predicted performance improvements \cite{stadelmann2009unfolding} from exploiting SST are still to be harvested, this holds large prospects. It could be realized, e.g., by combining a model that focuses on FBA with one regularized to focus on SST \cite{ganaie2022ensemble}. 

\begin{figure}[t]
  \centering
  \includegraphics[width=0.9\columnwidth]{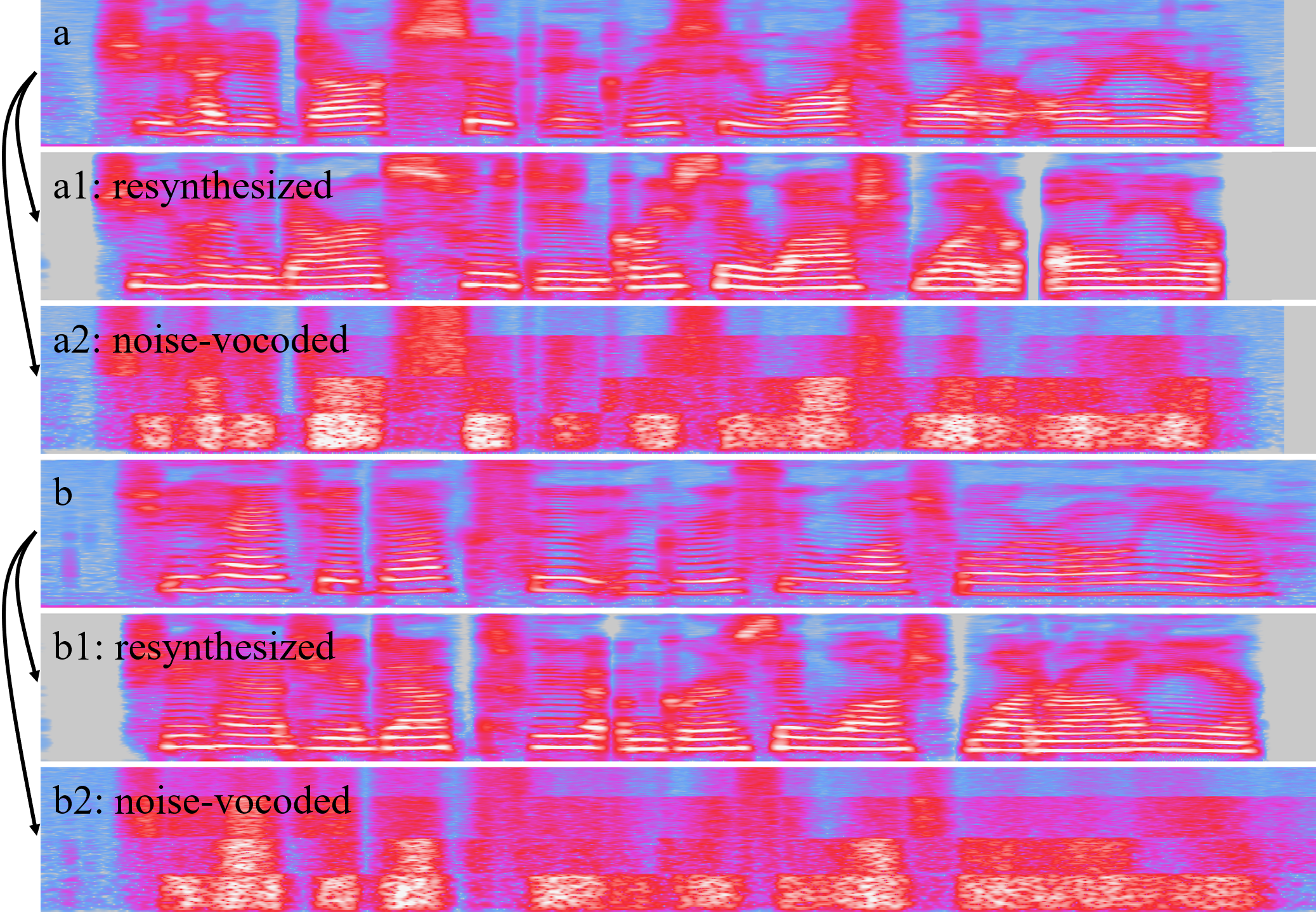}
  \caption{Visualization of FBA equalization: Compressed spectrograms of the same sentence (SA1) of a male (a: MDAB0) and female (b: FCJF0) TIMIT speaker with derived synthesized (a1/b1) and noise-vocoded (a2/b2) variants.}
  \label{fig_fba_equalization}
\end{figure}

\subsection{Methodology and experimental setup}

First, we test the conjecture that DNNs only model those features necessary to solve the task, starting with the easiest to learn. Therefor, we increase the difficulty of the task by including an acoustically more challenging dataset and check for an increase of actual SST modeling via the test established before, using the VoxCeleb corpus while keeping other parameters equal to the TIMIT analysis in Sec. \ref{sec_quantify} (i.e., for the CNN, RNN, and ResNet models we keep the experimental setup compatible with \cite{stadelmann2018capturing} while for the F-ResNet model we keep it compatible with \cite{chung2020in}). VoxCeleb1 contains $148,642$ utterances of varying length (few seconds up to several minutes) from $1,211$ different speakers recorded in the wild, thus holding a variety of background noises including cross talk; VoxCeleb2 contains more than $1$ million utterances from $5,994$ speakers. We use the standard experimental protocol of Chung et al. to train a SV system on VoxCeleb2 
, and evaluate SV performance on the ``hard'' test set of VoxCeleb1 \cite{chung2020in}. We omit SC results on VoxCeleb as experiments in Sec. \ref{sec_quantify} led to similar conclusions for both tasks.

Second, we aim at actively nudging models to learn SST by making FBA less attractive, inspired by the previous approach that scrambled SST. While scrambling the frequency axis by na\"ively randomizing the rows in a spectrogram would effectively remove SST as well (as it depends on the evolution of sound over frequency band borders), we instead propose the regularization strategy of largely \emph{equalizing} (instead of randomizing) FBA amongst speakers while retaining their discriminative SST. Note that this is done specifically to study the effect of devalued FBA on SST-modeling by our DNNs, not as a general way to augment speech data \cite{park19e_interspeech}. Specifically, we chose two equalization strategies (cf. Fig. \ref{fig_fba_equalization}) on the TIMIT corpus: (a) Noise-vocoding the sentences with just $4$ broad frequency bands \cite{shannon1995speech}, modulated by the original speakers' energy levels, to \emph{reduce} spectral differences; and (b) using a custom version of the Slang TTS \cite{slowsoft2021} speech synthesizer (able to use speaker-specific phoneme-level timing annotations and energy contours from the TIMIT corpus as additional input) to recreate every sentence from its text transcript with an identical synthetic voice, thus \emph{eliminating} spectral differences but keeping some SST.

\subsection{Results and discussion}

SV results on VoxCeleb are shown in the left part of Tab. \ref{tab_forcing}. Comparing them with SV on standard TIMIT (Tab. \ref{tab_timit_verification}) and using the cell annotations marked by $\Downarrow$, the following is noteworthy: (g) On TIMIT, competitive results are achieved using \ShuffleUtterance/\ShuffleUtterance\ (CNN) and  \ShuffleSeg/\ShuffleSeg\ (RNN, ResNet) that are only marginally worse than the best result per model and in particular strong in comparison with \OrigSeg/\OrigSeg. On VoxCeleb, \OrigSeg/\OrigSeg\ now clearly is better than all other combinations that include randomization by a larger margin. While the absolute scores are bad for the two simpler models, the ResNet shows a reasonable EER using \OrigSeg/\OrigSeg\ on VoxCeleb, only about $3$ times worse than the best result on the much simpler TIMIT and about twice as good as the best result involving any randomization. The F-ResNet shows state-of-the-art performance under \OrigSeg/\OrigSeg\ (i.e., normal) conditions, thereby conforming that the experimental setup and codebase used for these experiments is sound. We conclude that providing a more challenging task (one that cannot be solved relying on simpler frequency-domain features alone, as research prior to the deep learning era has shown \cite{stadelmann2010voice}) stimulates the exploitation of SST in models to some degree, depending on the ability of the model -- as all conditions with scrambled SST fall far behind in performance (by a factor of $\geq 2.75$ for the best model, F-ResNet).

The middle part of Tab. \ref{tab_forcing} contains the results on noise-vocoded TIMIT, where speakers' individual timbre has been largely removed. Again comparing with the respective results on standard TIMIT, it is noteworthy that (h) best results are now achieved by \OrigSeg/\OrigSeg\ for the RNN and ResNet. However, the CNN still has best results involving random timing, and also for the RNN and ResNet models, the margin for \OrigSeg/\OrigSeg\ is small and EERs are $4$-$5$ times higher than on standard TIMIT. We conclude that the effect seen on VoxCeleb (a harder task makes the models start learning SST) is visible to some degree, but less pronounced. (i) The same is true for the results on resynthesized TIMIT (cf. right part of Tab. \ref{tab_forcing}), except that best results for the CNN are achieved using another form of randomization and the EERs are $8$-$11$ times worse here. (k) It is different for the F-ResNet, where best results on noise-vocoded TIMIT are still achieved using \ShuffleSeg/\ShuffleSeg\ (and only with a very thin margin using \OrigSeg/\OrigSeg\ on resynthesized TIMIT). We conjecture that this is again due to the tininess of the TIMIT database for training this large model (see (j) discussed in Sec. \ref{sec_quantify}). 

We conclude that diminishing the dominance (i.e., speaker discriminativeness) of FBA partially (using noise vocoding) or fully (through resynthesis) brings forth some exploitation of SST for SV, but not optimally (which would be evidenced not necessarily by lower absolute EERs, but by larger margins between top \OrigSeg/\OrigSeg\ results and everything else). Moreover, such an effect is even less pronounced when switching tasks from SV to SC (tables omitted): No benefit of \OrigSeg/\OrigSeg\ can be observed on TIMIT-NV and only little evidence for it is seen on TIMIT-Syn. We conjecture that the attempt to model SST apart from FBE is suboptimal (in accordance with the literature \cite{rose2002forensic, hansen2015speaker} that categorizes SST as of subordinate importance but helpful in addition to FBA). Evidence for this is the experiment on VoxCeleb that shows that under challenging circumstances ResNets can achieve unparalleled results by exploiting FBA and SST jointly.

\section{Conclusions, future work and limitations}
\label{sec_conclusion}

In this paper, we have presented the first systematic study on learning supra-segmental temporal features by DNNs for SR. Not focusing on presenting a new kind of model or SR methodology, we have instead shown that state-of-the-art CNN, RNN and ResNet models for SV and SC, when trained on clean data, simply ignore any useful supra-segmental temporal cues in the audio signal despite contrary conjectures in the literature and the models' principal abilities to learn such features. We have called this phenomenon ``deep cheating''. It is relevant since related work provided evidence \cite{stadelmann2009unfolding} that improved modeling of such higher-level features should result in one order of magnitude lower error rates in related tasks and hence holds a key for targeted future research, guided by our test to quantify actual SST explotation. It is also of importance in the context of explaining \emph{how} DNNs achieve their superior results (XAI), where our explanation goes beyond activation visualization to explain individual classifications 
towards a broader understanding of signal processing by DNNs. 

Furthermore, we have presented two approaches to force DNNs to exploit SST, and measured their effectiveness: (a) Increasing task difficulty by using acoustically more challenging data (VoxCeleb instead of TIMIT), and (b) removing the discriminative power of FBA by equalizing speakers' timbre. The results indicate that both approaches achieve respective results nominally, thereby confirming other studies that attest DNNs laziness in modeling only the easiest available features to solve a given task. Theoretical 
and empirical studies suggest that scaling up training time might help overcome such imperfect local minima \cite{
power2022grokking}.

We have conducted extensive experiments to verify the correctness and stability of our results for a wide range of design choices. Our claims hold for reasonable settings of the hyperparameters learning rate, number of epochs, segment and hop length, and embedding size; when using the original loss functions of published models instead of speed-ups; with varying number of frequency bands for the noise vocoder or using MBROLA \cite{dutoit1996mbrola} as another synthesizer; and for evaluating on VoxCeleb2. TIMIT, though small, is a sound basis for our findings (cf. \cite{power2022grokking}): It has been used successfully for this purpose by the community before; we do not observe problems with overfitting for all but the F-ResNet model, although by construction of the mini batches, we only exploit a fraction of the available training data; it contains pure voices without exogenous difficulties (noise, brevity, \dots), offering to study SR capability in isolation and hence granting an unbiased look at DNNs' abilities for voice modeling.

However, we have also shown that attempting to learn SST apart from FBA results in severely underperforming models that verge on random SV results and are even less helpful for SC. Hence, our results are preliminary with respect to finding better ways of exploiting speaker-specific SST with DNNs. Thinking of perceivable instantiations of SST like personal linguistic melody, it is elusive how such strong rhytmic-prosodic patterns are not picked up by any of the most capable general pattern recognition methods we know today, deep neural networks. Future work should therefor concentrate on finding inductive biases for deep networks that fully exploit the time axis for speaker specificity. Inspired by how auto-regressive self-supervised learning in text processing works \cite{mikolov2013efficient} and building on the success of respective large language models, transformer-based architectures and large-scale pre-training could be a way to integrate handling of dynamic features either directly into the model or into methods for speech augmentation. Such work should use the test presented in Sec. \ref{sec_quantify} to benchmark the success of actually exploiting STT. Our code is available online at \url{https://tinyurl.com/deepcheating}. 

\section*{Acknowledgments}

The authors thank Bernd Freisleben, Claude Olivier Lehmann, Frank-Peter Schilling and Richard Hahnloser for encouraging discussions, and their collaborators on voice recognition methodology over the years: Oliver D\"urr, Yanick X. Lukic, Carlo Vogt, 
Feliks Hibraj, Sebastiano Vascon, Marcello Pelillo, 
Patrick Gerber, 
Sebastian Glinski-Haefeli, 
Benjamin Neusser, Savin Niederer, 
Timon Gygax, J\"org Egli, 
Niclas Simmler, Amin Trabi, 
Jan Sonderegger, Patrick Walter, 
Christian Lauener, 
Pascal Sager, 
and Sydney M. Nguyen. 

\bibliographystyle{elsarticle-num}
\bibliography{main}


\end{document}